\documentstyle[11pt,newpasp,twoside,epsf]{article}

\begin{document}

\title{Mining for Metals in the Ly$\alpha$ Forest}
\author{Sara L. Ellison}
\affil{Institute of Astronomy, Cambridge, UK}
\author{Antoinette Songaila}
\affil{Institute for Astronomy, University of Hawaii, USA}
\author{Joop Schaye, Max Pettini}
\affil{Institute of Astronomy, Cambridge, UK}

\begin{abstract}
In order to ascertain the extent of metal enrichment in the Ly$\alpha$
forest, we have analysed a very high S/N spectrum of the $z = 3.625$ QSO
Q1422+231.  We find that in high column density Ly$\alpha$ clouds, the
power law column density distribution function of C~IV continues down to log
$N$(C~IV) = 11.7.  In addition, by analysing pixel-by-pixel optical
depths we show that there are considerably more metals in the
Ly$\alpha$ forest than are currently directly detectable.
\end{abstract}
\vspace{-1cm}
\section{Introduction}

Our understanding of the Ly$\alpha$ forest and its connection with the
Intergalactic Medium (IGM) have undergone radical revision in recent
years.  The paradigm in which the IGM consists of discrete, isolated
clouds has been revolutionised by a generation of hydrodynamical
simulations (e.g. Hernquist et al 1996).  These simulations show
that the `bottom-up' hierarchy of structure formation knits a
complex but smoothly fluctuating cosmic web, consisting of filaments,
knots and extensive `voids'.  Originally thought to
be chemically pristine, it is now well-established that a large
fraction of the strongest Ly$\alpha$ absorbers exhibit some metal
enrichment, most notably C~IV (e.g. Cowie et al 1995).  
Here, we address two specific
questions.  Firstly, are the C~IV absorbers that have thus far been
detected just the tip of the iceberg and can more sensitive spectra
mine ever weaker systems?  Secondly, to what H~I column densities does
the enrichment extend?  This latter point has particularly poignant
implications for the origin and transport mechanism of these metals.
In-situ formation in a nearby galaxy could be responsible for the
enrichment of its local IGM, explaining the presence of C~IV
in the relatively high column density Ly$\alpha$ clouds.  However, 
low column density Ly$\alpha$ clouds (log$N$(H~I) $<$ 14.0), which
are associated with physically less dense regions, are found 
further from the sites of
star formation.  The presence of C~IV in these clouds could be
indicative of widespread metal enrichment, possibly by an early
epoch of Population III star formation.  This is an overview of work
described in more detail in Ellison et al (2000).

\section{C~IV in High Column Density Ly$\alpha$ clouds}

\begin{figure}
\plotfiddle{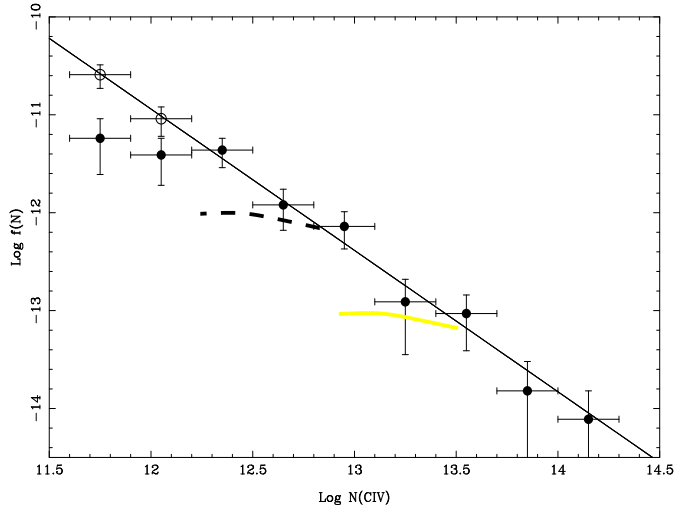}{6.5cm}{270}{40}{40}{-150}{220} 
\caption{\label{f_N} Column density distribution of C~IV absorbers in
Q1422+231. Points corrected for incompleteness are shown with open
circles.  The approximate turnovers seen in 
previous determinations of the $f(N)$ power law are shown as a grey
solid line (Petitjean \& Bergeron 1994) and black dashed line
(Songaila 1997).} 
\end{figure}

With a S/N ratio in excess of 200 redward of Ly$\alpha$, our Keck/HIRES
spectrum of Q1422+231 is one of the most sensitive currently
available to search for C~IV associated with the Ly$\alpha$ forest.
Although this target has been extensively studied in the past, our
data reveal several C~IV systems that had not been previously detected
in spectra of lower S/N.  We undertake the standard procedure of
fitting Voigt profiles to determine column densities, $b$-values and
redshifts of the 34 detected C~IV systems associated mainly
with Ly$\alpha$ clouds with log $N$(H~I) $>$ 14.5.  Previous studies
of C~IV absorbers have established a power law column density
distribution of the form $f(N)dN = BN^{-\alpha} dN$
and determined $\alpha \sim 1.5$, complete down to log $N$(C~IV)
$\simeq$ 12.75 for $z > 3$ (Songaila 1997).  Below this limit, there is an
apparent departure from the power law which could be due to either
incompleteness or a real turnover in the number density of C~IV systems.
We perform a maximum likelihood fit to our data points and determine a
power law index $\alpha = 1.44 \pm 0.05$, in good agreement with
previous estimates (see Figure \ref{f_N}). 
The column density limits at which previous studies have
exhibited a departure from the power law are sketched and clearly our
data establish that this was due to incompleteness.  The C~IV systems
from this single high quality spectrum are sufficient to show that
the power law continues down to at least log $N$(C~IV) $\sim$
12.3, below which the data points start to turn-over.  Again, this could be due
to the incompleteness caused, for example, by a bias against weak C~IV
lines with large 
$b$-values, or evidence of a real turn-over in column density distribution.
By simulating C~IV lines for the two lowest column
density intervals in Figure \ref{f_N} with $b$-values drawn at random
from the observed distribution, we can estimate the incompleteness
correction factor by determining the frequency with which these lines
are recovered.  Once this has been taken into account, there is no
shortfall compared with the power law, showing that it
continues at least down to log $N$(C~IV) = 11.7, a factor
of ten more sensitive than previous analyses.

\section{Probing the Low Column Density Ly$\alpha$ Forest}

\begin{figure}
\plotfiddle{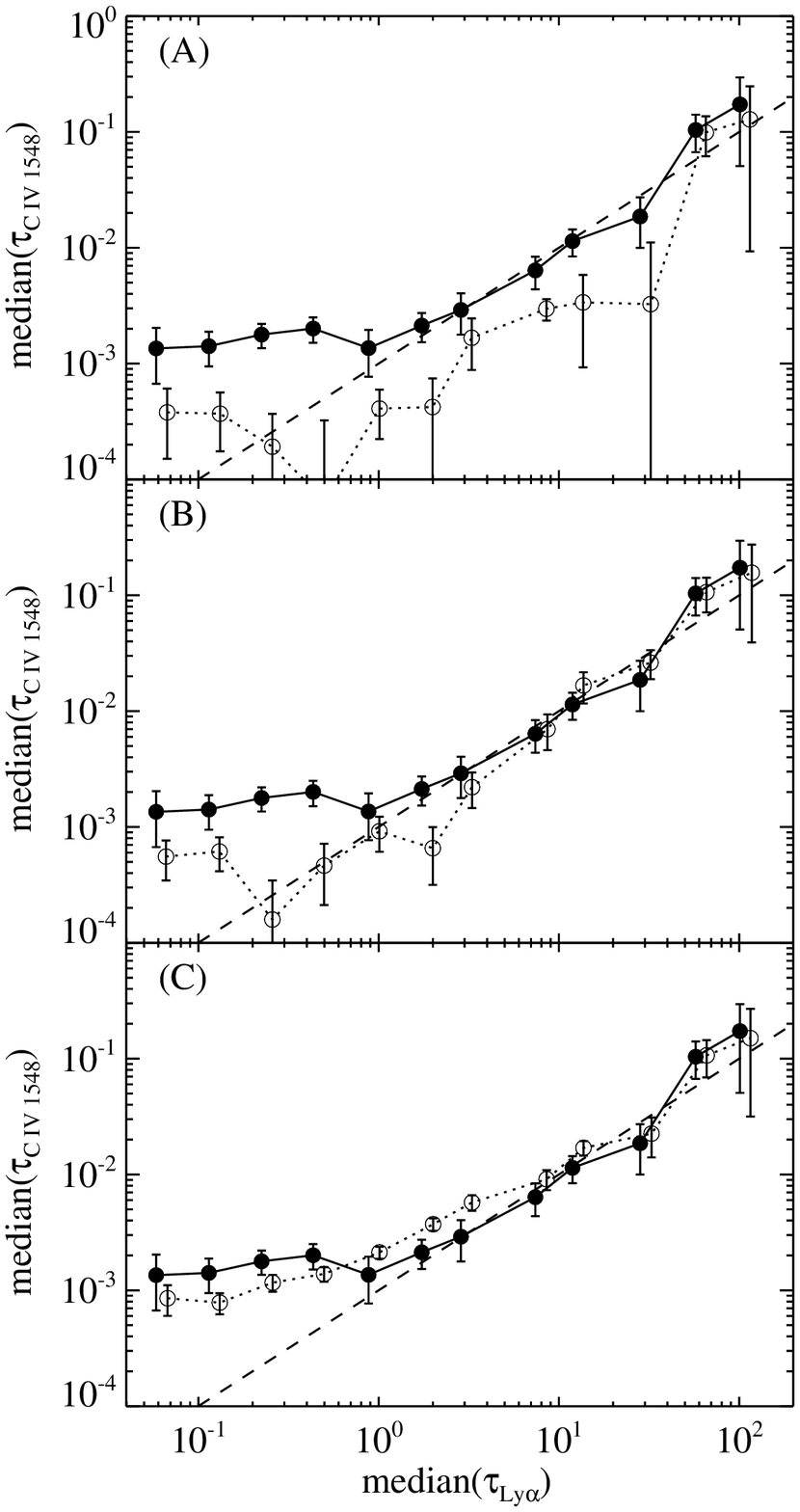}{10cm}{0}{48}{48}{-90}{-20} 
\caption{The results from the optical depth analysis of Q1422+231
(solid points) are compared with three synthetic spectra.  Top panel:
open circles show the measured optical depths in a synthetic spectrum
enriched solely with the detected C~IV systems.  
Middle panel: In addition to the detected C~IV
systems, log $N$(C~IV)= 12.0 is included in all Ly$\alpha$ clouds with log
$N$(H~I) $> 14.5$.  Bottom panel:  Supplementary C~IV is now added in
all weak (log $N$(H~I)$<14.5$) Ly$\alpha$ lines with log C~IV/H~I =
$-2.6$.} 
\label{testabc}
\end{figure}

Direct detection of the C~IV systems associated with low column
density Ly$\alpha$ clouds (log $N$(H~I) $<$ 14.0) is observationally
very challenging due to the extreme weakness of the absorption.
In the past, efforts have been made to overcome this problem by
stacking together the regions where C~IV absorption is expected in
order to produce a high S/N composite spectrum (e.g Lu et al 1998).  In Ellison
et al (1999), we showed how a random redshift offset between the
Ly$\alpha$ line and its associated C~IV feature could `smear' out the
stacked feature and consequently underestimate the amount of metals
present. Instead, we favour the optical depth method developed by
Cowie \& Songaila (1998) which we have found is more robust against
redshift offset (see Ellison et al 2000 for a detailed
discussion of this point).  Briefly, the optical depth method consists
of stepping through the spectrum and measuring the optical depth
($\tau$) of each Ly$\alpha$ pixel and its corresponding C~IV.  The
results of this analysis are shown by the solid line in Figure
\ref{testabc} and are consistent with a constant  level of C~IV/H~I
(as shown by the dashed line) for optical depths down from
$\tau$(Ly$\alpha$) $\sim 100$ over two orders of magnitude, below
which $\tau$(C~IV) flattens off to an approximately constant value.
In order to interpret these results, simulated spectra were reproduced
using the Ly$\alpha$ forest taken directly from the data, adding C~IV
with a given enrichment recipe and then analysing the synthetic
spectrum with the optical depth technique.  A total of 3 simulated
spectra were created, all of which include a random redshift offset of
17 kms$^{-1}$ between Ly$\alpha$ and C~IV and assume $b$(C~IV) = 1/2
$b$(Ly$\alpha$).  Spectrum `A' includes only the directly detected C~IV 
with no additional enrichment and the optical
depth analysis reveals that it is clearly C~IV deficient in comparison
with the data at almost all optical depths.  Clearly, there is more
C~IV in the data than is accounted for in the 34 directly identifiable
systems.  Adding C~IV to the log
$N$(C~IV) $>$ 14.5 Ly$\alpha$ clouds at the detection limit of the
spectrum (log $N$(C~IV) = 12.0) can reproduce the results obtained for
the data for $\tau$(Ly$\alpha$) $>$3 but not at lower optical depths
(spectrum `B'). The data are more consistent with spectrum `C' in
which a constant C~IV/H~I ratio of $-2.6$ was also included in log $N$(H~I)
$<$ 14.5.  

We investigated the possible limitations of our analysis due to such
effects as contamination by other metal lines, errors in the continuum
fit and scatter in the C~IV/H~I ratio.  We conclude that overall this
is a robust technique and that the limiting factor is likely to
be the accuracy of the continuum fit in the Ly$\alpha$ forest
regions which could mimic the flattening of $\tau$(C~IV) that we
observe in the data  at low H~I optical depths.  Nevertheless, we find
that even in the high optical depth H~I pixels (that will not be
seriously affected by small continuum errors) the 
identified C~IV systems are insufficient to account for the all the
measured absorption and that there are clearly more metals in the IGM
than we can currently detect.

\acknowledgements

SLE is very grateful to the LOC for their generous financial
assistance towards attending this conference.

\end{document}